# Recommended Actions for the American Astronomical Society: CSWA's Perspective on Steps for a more Inclusive Astronomy - II. Recommendations

Rachel Wexler[1] Patricia Knezek[1] Gregory Rudnick[2] Nicolle Zellner[3] Kathleen Eckert[1] JoEllen McBride[4] Maria Patterson[1] Christina Richey[5] Committee on the Status of Women in Astronomy

[1]Committee on the Status of Women in Astronomy, [2]University of Kansas, [3]Albion College, [4]University of Pennsylvania, [5]Jet Propulsion Laboratory/California Institute of Technology

**ABSTRACT**

This paper, the second in a series of two, provides a set of recommendations that the American Astronomical Society (AAS) can take to create a more diverse and inclusive professional society for astronomers, with a focus on women astronomers. As noted in Paper I, now is the time for the AAS to take decisive action to transform astronomy into a diverse and inclusive profession. By combining the results of our 2019 survey, which is described in Paper I, peer-reviewed academic literature, and findings from many of the white papers submitted to Astro2020, the CSWA has developed 26 specific actions the AAS can take to help end harassment and bullying in astronomy; advance career development for astronomers who are women, members of other underrepresented groups, and intersections of these populations; and improve the climate and culture of AAS meetings. Actions to reduce rates of harassment and bullying include improvements to the AAS's anti-harassment policies and procedures and the development of astronomy-specific anti-harassment training resources. Actions to advance career development include creating a compensation database, improving how jobs are posted in the AAS Job Register, and supporting/enhancing a distance mentorship program. Finally, we call on the AAS to continue improving the accessibility of AAS meetings and to continue to support meeting sessions whose focus is to discuss issues of equity, diversity, and inclusion.

## 1. Introduction and Summary of Recommendations

As described in Paper I, the American Astronomical Society's (AAS) Committee for the Status of Women in Astronomy (CSWA) was created in 1979 with the mandate to review the status of women in astronomy and to make practical recommendations to the AAS Council (now the AAS Board of Trustees) on actions to advance the status of women in astronomy. In the following, the CSWA interprets "women" to mean people who identify as female, including trans women, genderqueer women, and non-binary people who are significantly female-identified (Knezek et al., 2020) . This series of two papers represents the committee's effort to communicate the voice of the AAS community to the AAS about the current priorities the AAS should focus on that are particularly relevant to women in the field.

In this series of two papers, the CSWA draws upon its 2019 community survey (Zellner et al., 2024), which formed the foundation for the CSWA's Strategic Plan for the 2020s, and lays out the objectives that will guide the CSWA's activities for the next decade. Paper I presents the background and methods used to derive our recommendations, while Paper II (this paper) presents recommendations that align with both the CSWA's Strategic Plan and Priorities 2 and 3 of the AAS's 2021-2026 Strategic Plan (AAS, 2021). All of the recommendations support the CSWA's progress towards its mission to build an inclusive and self-sustaining community that supports gender equity and the success of women in astronomy.

This paper is organized as follows. In Section 2, we advocate for the need for an Office of Diversity and Inclusion within the AAS organizational structure. In Section 3, we present recommendations to eliminate harassment and bullying: Regularly Update Internal Anti-Harassment and Bullying Policies and Procedures (Section 3.1), Address Minor Forms of Harassment and Bullying Using a Restorative Approach (Section 3.2), Create an Online Guide for Targets of Harassment and Bullying (Section 3.3), Employ an Advocate to Support Targets of Harassment and Bullying (Section 3.4), and Create, Administer, and Support Astronomy-Specific Anti-Harassment and Bullying Training (Section 3.5). In Section 4, we present recommendations to advance career development: Partner with Social Scientists to Research the State of the Profession (Section 4.1), Model Equitable Selection Processes for Committee Membership and Awards (Section 4.2), Improve AAS Journal Review Processes by Requiring Dual-Anonymous Review and by Releasing Article Submission and Acceptance Data (Section 4.3), Develop a Compensation Database with a Focus on Graduate and Postdoctoral Compensation (Section 4.4), Take Steps to Improve the Graduate and Postdoctoral Experience (Section 4.5), Address the Two-Body Problem (Section 4.6), Continue to Educate Astronomers About Alternative Careers (Section 4.7), Support a Distance Mentorship Program (Section 4.8), and Advocate for a Better Profession (Section 4.9). In Section 5, we make recommendations to continuously improve AAS meetings: Continue and Improve Efforts to Advance Inclusivity at AAS Meetings (Section 5.1) and Continue to Support Diversity and Inclusion Sessions at AAS Meetings (Section 5.2). We summarize in Section 6. In the Appendix we present our quantitative data for the Likert-opinion scale questions. The free-response replies are provided in an on-line repository(Zellner et al., 2022)  and are summarized in Paper I.

*In sections 2-5, we provide a comprehensive set of actions that the AAS should take within the next decade to reduce harassment and bullying and to advance diversity and inclusion in astronomy.* Table 1 provides a summary of the main recommendations in this paper, referencing the appropriate section of the paper in which the recommendation is made. Many of these recommendations have sub-recommendations that provide more detail. The full list of all recommendations and sub-recommendations can be found in Appendix A.

The recommendations in this paper are not intended to be prescriptive. Due to the complex issues surrounding the implementation of some of the recommendations, we intend for the AAS governance to decide on details of prioritization, implementation, and assessment of impact.

**Table 1.** Summary of recommendations. Details are provided in sections 2-5.

| Table 1 | |
|---|---|
| Section | Brief recommendation text |
| 2 | The AAS should establish an Office of Diversity and Inclusion (ODI) with two to three staff members, as a division of the Executive Officer's staff. |
| 3.1 | The AAS should commit to critically assessing and updating its definitions of harassment and bullying, and its harassment and bullying reporting procedures, investigative processes, and response policies regularly (e.g., at least every five years). |
| 3.1 | AAS staff members should conduct a full review of literature on harassment and bullying in the sciences and best practices for addressing reports of harassment and bullying, with a focus on materials published in the past five years. |
| 3.1 | The AAS should collect the input of its membership regarding the effectiveness of its anti-harassment and anti-bullying policies. |
| 3.1.1 | The AAS should ensure consistency in how harassment and bullying are defined in the AAS's Code of Ethics and the Anti-Harassment Policy for AAS & Division Meetings & Activities. |
| 3.1.2 | The AAS should make changes to the content and procedures contained in the sections of the Anti-Harassment Policy for AAS & Division Meetings titled "Reporting an Incident" and "The Investigation". |
| 3.2 | The AAS, preferably through the ODI introduced in Section 2, should create and manage an information escrow to be used to identify perpetrators of minor forms of bullying and harassment. Information escrow is a process by which parties can pass on information to a neutral third party. |
| 3.3 | The AAS should create a detailed guide to navigating and reporting harassment and bullying and feature it on the AAS website. |
| 3.4 | The AAS should employ a trained, experienced advocate to work as a member of the Executive Officer's staff, preferably within an ODI. |
| 3.5 | The AAS should develop astronomy-specific anti-harassment and anti-bullying trainings, administer them during meetings, and share astronomy-specific anti-harassment and anti-bullying training materials among institutions. |
| 3.5.1 | The AAS should require all members and meeting attendees, including regional and division meeting attendees, to complete a short online interactive training (e.g., a video that addresses multiple types of harassment and bullying, as well as mental health, physical health, and intellectual impacts of harassment and bullying) before registration and require attendees to answer a small number of questions regarding common misconceptions about harassment and bullying. |
| 3.5.2 | The AAS should publish, maintain, and publicize a list of vetted training programs that are relevant to issues facing Astronomy & Physics departments. |
| 4.1 | The AAS ODI, in coordination with the diversity committees and the Demographics Committee, should support and expand partnerships with social scientists and other experts to conduct studies that track and examine the factors that affect astronomers' career paths into the 2020s and beyond. |
| 4.2 | The AAS should evaluate and revise its nomination and selection processes for committee membership and awards every two years using a similar process to the framework detailed in Section 3.1 for reviewing and revising anti-harassment and anti-bullying policies and procedures. |
| 4.3.1 | The AAS journals should lead the physical sciences in implementing dual-anonymous review for journal articles. |
| 4.3.2 | The AAS should systematically collect and publish article submission and acceptance data for its journals, including the rates at which each journal receives and accepts articles from women and members of historically marginalized groups compared to men and white authors. |
| 4.4 | The AAS should create an official compensation database and prioritize gathering data on compensation for graduate students and postdoctoral researchers. |
| 4.5 | The AAS should host sessions at AAS meetings for astronomers to convene to discuss the challenges of early-career employment. |
| 4.5 | The AAS should encourage grant PIs to offer paid leave (within institutional guidelines) for all graduate students and postdocs. |
| 4.6 | The AAS should facilitate dialogue on the two-body problem at AAS meetings and sponsored events. |
| 4.6 | The AAS should encourage its membership to implement the lessons learned from remote work during the COVID-19 global pandemic to integrate more remote work into their strategies for assisting dual-career couples in the long-term. |
| 4.6 | The AAS should create and distribute a best-practices guide on the two-body problem. |
| 4.7 | The AAS should work with both academic and non-academic partners to create and promote more educational materials about career alternatives. |
| 4.8 | The AAS should work with professionals with expertise in mentoring within the physical sciences to develop an effective mentoring program and to recruit astronomers from both academic and non-academic institutions to serve as mentors. |
| 5.1 | AAS should continue to provide at least some online access to selected aspects of its meetings. |
| 5.2 | The AAS should take care to avoid scheduling sessions on diversity, equity, and inclusion at overlapping times. |

## 2. Recommendation for The Need for an Office of Diversity and Inclusion

To coordinate actions to end harassment and bullying and advance diversity and inclusion in astronomy, we recommend that the AAS establish an Office of Diversity and Inclusion (ODI) with two to three staff members, as a division of the Executive Officer's staff. The American Physical Society has a director for diversity and education, and the AGU employs two staff members to manage scientific ethics (AGU considers harassment and bullying violations of scientific ethics). The AAS has a very small staff compared to these larger professional societies, but it and its membership would nonetheless benefit from the addition of a few people who are specifically dedicated to working on initiatives related to ending harassment and bullying and increasing diversity and inclusion (American Geophysical Union, n.d.; American Physical Society, n.d.).

The diversity/inclusion committees (CSWA; Committee on the Status of Minorities in Astronomy, CSMA; Committee for Sexual-Orientation & Gender Minorities in Astronomy, SGMA; and Working Group on Accessibility and Disability, WGAD) are made up of active astronomers who volunteer their time to work on committee initiatives and projects, but they cannot, and should not be expected to, perform the day-to-day work that is required to develop, implement, and assess the initiatives that we recommend the AAS undertake. This work should be done by paid professionals and experts. Many of our survey respondents reported that diversity initiatives often pressure women and scientists from historically marginalized groups[1] to be *the* experts on every issue and to dedicate their limited free time to diversity work. They acknowledged that diversity experts and anti-harassment and anti-bullying professionals exist, and requested that the AAS consult with and employ those professionals. These committee members work to fulfill committee mandates to make practical recommendations to the AAS, but without dedicated staff at the AAS to take on the responsibility of carrying out recommendations, the impact is limited. Many recommendations are time and resource intensive. For these initiatives to reach their full potential, they require not only resources, but also attention, care, personal commitment, and assessment by people who are passionate about transforming astronomy into a leading scientific field for diversity and inclusion.

To support the core ODI staff, we recommend that the AAS also consider funding internships for aspiring social scientists and astronomers to work with the ODI and the diversity/inclusion committees on some of the initiatives and actions described in this paper, though not all of our recommendations are appropriate intern projects. However, an intern could assist with recommendation 3.3, on creating an anti-harassment and anti-bullying training video to facilitate a shared understanding of harassment and bullying.

Creating an ODI would demonstrate a serious commitment on behalf of the AAS to transforming astronomy for the better. An ODI would play an important and impactful role by directly carrying out diversity and inclusion initiatives, and helping coordinate the activities of the diversity/inclusion committees and other committees involved in inclusion-related activities (such as the Site Visit Oversight Committee, the Education Committee, and others). Throughout these recommendations, we refer to responsibilities that should be delegated to AAS staff, and our preference is that these staff members are ODI staff.

Recently, the AAS hired a DEI (Diversity, Equity, and Inclusion) Committee Support Specialist, whose job description included aiding the Board of Trustees and the four diversity/inclusion committees (CSWA, CSMA, WGAD, SGMA) "with DEI issues and positively and proactively addressing the issues from our growing diverse community" (AAS job ad, posted 4/18/2022). That position was staffed for a short period of time and was beneficial while filled, especially in facilitating the coordination of the different diversity committees, which in turn is instrumental in achieving an intersectional approach to DEI. We advocate for the prompt creation of an official ODI, in which DEI Committee Support presumably will play an important role.

# 3. Recommendations to Eliminate Harassment and Bullying

Harassment and bullying of women is common in astronomy (e.g., [Clancy et al. 2017](#)). We take the stance that the AAS can and should be a leader in the effort to end harassment and bullying in astronomy and set an example for other scientific professional societies[2]. As noted in the report from the National Academies of Science, Engineering, and Mathematics [(NASEM, 2018)](#), professional societies have the potential to be powerful drivers of change through their capacity to help educate, train, codify, and reinforce cultural expectations for their respective scientific, engineering, and medical communities. Through a commitment to reducing harassment and bullying, put into action through improvements to its policies, the AAS can create a more inclusive profession, thereby boosting the retention of women and members of historically marginalized groups in astronomy. The AAS must make a stronger effort to communicate its policies, to engage more astronomers in anti-harassment and anti-bullying training, and to support those who have been targets of harassment and/or bullying. The AAS can best support the development and implementation of these recommendations by consulting with and hiring anti-harassment and anti-bullying professionals in any such process, potentially via the inclusion of individuals with such expertise in a new ODI (See [Section 2](#).)

## 3.1 Commit to Regularly Updating Internal Anti-harassment and Anti-bullying Policies and Procedures

We recommend that the AAS should commit to critically assessing and updating its definitions of harassment and bullying, and its harassment and bullying reporting procedures, investigative processes, and response policies regularly (e.g., at least every five years). Harassment and bullying can be sexual, and it also may be racial, anti-LGBTQ+, ageist, or a combination of these, or it may not fit any common category. It is important for the AAS to specifically define harassment and bullying and to include in its policies specific examples for common types of both of these. Definitions that are comprehensive and clear will provide a basis for better and more widespread anti-harassment and anti-bullying education (see Sections [3.3](#) and [3.5](#)).

Scheduled, systematic updates will ensure that the AAS's definitions of harassment and bullying continue to be inclusive of the prevalent forms of harassment and bullying that astronomers face, especially as new forms of harassment and bullying surface and older forms are revised. The AAS's definitions and policies must evolve to be relevant to an evolving cultural landscape in order to serve as examples that astronomical institutions and departments can follow.

These updates will also improve the monitoring and enforcement of anti-harassment and anti-bullying policies by creating space for continuous assessment of what is working and what is not. Monitoring and enforcement are critical. Of our survey respondents, 83% rated enforcing anti-harassment and anti-bullying policies as a somewhat effective or very effective policy that the AAS could pursue to prevent harassment and bullying. Policies and procedures are only as effective as their implementation.

The AAS last updated its anti-harassment and anti-bullying policy for meetings in October of 2015, according to the date in the title of the file posted on the AAS web page (Anti-Harassment Policy for AAS & Division Meetings & Activities, 2017). Over the past five years, our understanding of the forms and impacts of harassment and bullying in the sciences have shifted, partially due to the publication of the 2018 NASEM report. Further, the definition of sexual harassment contained within the AAS Code of Ethics does not match the definition of harassment in the Anti-Harassment Policy for AAS & Division Meetings & Activities. These definitions must be amended to be consistent with one another.

In the following paragraphs, we outline a framework for performing community-informed policy updates. We propose this framework for anti-harassment and anti-bullying policies and procedures, an area we consider to be most critical, but it may be applied to any area of policy.

Updating the anti-harassment and anti-bullying policies and procedures should begin with two sources of input. First, we recommend that AAS staff members should conduct a full review of literature on harassment and bullying in the sciences and best practices for addressing reports of harassment and bullying, with a focus on materials published in the past five years. A full review should rely on peer-reviewed publications, but should also include policy white papers published by astronomers and groups of astronomers (like this one and many others submitted to Astro2020 and the Planetary 2023 decadal survey), findings from surveys and studies conducted by astronomical institutions (like the Maunakea report; Dempsey, 2018), and materials published by professional societies that serve similar communities, including the American Geophysical Union (AGU) and the American Physical Society (APS). A full review may examine best practices for addressing harassment and bullying used by universities, companies, and other organizations. The AAS may also hire an outside consultant to provide input on its policies and procedures.

Second, we recommend that the AAS should collect the input of its membership. AAS staff should distribute a survey to assess the effectiveness of its anti-harassment and anti-bullying policies, and should reach out to department heads at universities, directors of telescopes, and other community leaders to ask them to submit input and to encourage their communities to participate in the survey. To accommodate the preferences of all potential participants, survey respondents should be given the option to submit anonymously or non-anonymously and the option to submit their contact information to allow the AAS to follow up with them, should they have ideas or questions. AAS staff should meet with members of each diversity committee (CSWA, WGAD, SGMA, and CSMA) and give each committee ample time to submit their own report of recommendations for the policy update.

Once each information gathering step has been completed, AAS staff should draft an update to the AAS's anti-harassment and anti-bullying policies and procedures, and circulate it to the diversity committees for review. Finally, once each diversity committee approves the update, the update should be submitted to the usual AAS policy approval processes.

### 3.1.1 Recommended Updates to the Current Definition of Harassment and Bullying

First, we recommend that the AAS ensure consistency in how harassment and bullying are defined in the AAS's Code of Ethics and the Anti-Harassment Policy for AAS & Division Meetings & Activities. Our recommendation for the structure and content of the AAS's definition of harassment and bullying is as follows.

We recommend that the AAS's unified definition of harassment and bullying should include a baseline definition of harassment and bullying followed by subsections for sexual, racial, ableist, ageist, and forms of harassment and bullying that may impact LGBTQ+ scientists, including harassment and bullying on the basis of sexuality, gender identity, and gender presentation. A comprehensive definition will clarify these forms of harassment and bullying with examples, but will also note that harassment and bullying may not fit into any one of the described categories. It should explain that those with intersectional identities may experience a multiplied effect of multiple forms of harassment and bullying.

We also recommend that the AAS's definition of sexual harassment should be updated to be consistent with the findings of the 2018 NASEM report, which identified three types of sexual harassment: sexual coercion, unwanted sexual attention, and gender harassment. Currently, the AAS Code of Ethics describes two types of sexual harassment: "quid pro quo harassment" and "hostile work environment." These categorizations are limited to the legal definition of sexual harassment and do not reflect the full reality of sexual harassment.

A comprehensive definition of sexual harassment should explain that repeated instances of gender harassment can lead to the same impacts on individuals as sexual coercion and unwanted sexual attention ([NASEM, 2018](); [Cedeno and Bohlen, 2022]()). It will also specify that harassment is assessed based on its impact on the target, and not the intention of the harasser [(NASEM, 2018)](). Finally, a complete definition will demonstrate that the AAS is aware of the full range of impacts harassment can have on scientists' mental and physical health *and* the ways harassment restricts scientific progress.

We acknowledge the prevalence of racial, anti-LGBTQ+, and ableist harassment and bullying in science, and call on the AAS to collaborate with the diversity committees to determine comprehensive definitions of each, and to include them as part of the AAS's policies.

### 3.1.2 Recommended Changes to Anti-harassment and Anti-bullying Policy Actions

In this section, we recommend changes to the content and procedures contained in the sections of the Anti-Harassment Policy for AAS & Division Meetings titled "Reporting an Incident" and "The Investigation" (Anti-Harassment Policy for AAS & Division Meetings & Activities, 2017). Our survey respondents overwhelmingly called for swifter and stronger action by the AAS when it receives reports of harassment and bullying. Notably, we received survey responses from astronomers who remarked that their cases were inexplicably dropped after they reported to the AAS.

We recommend that the AAS investigate *every* claim of harassment and bullying that it receives and treat cases of harassment and bullying as seriously as it treats other forms of ethical misconduct. Presently, all forms of ethical misconduct, including harassment and bullying are evaluated by a group of members of the Code of Ethics Committee (CoEC). All members of the CoEC who take part in investigations should be trained on investigating instances of harassment and bullying. In addition, we recommend that the AAS should engage with outside expertise to aid in the investigation(s) of instances of harassment and bullying. This is important as outside experts will undoubtedly be better trained than astronomers. Proper training is essential for effective investigations and will make the procedure for addressing harassment and bullying robust and fair.

The "Reporting an Incident" section of the Anti-Harassment Policy provides little information to support targets in their decision to report harassment and bullying. Presently, the only resource provided in this section is a link to the homepage of the CSWA website, which does not refer to the relevant section of the CSWA page and, in any case, was not designed to be a resource for targets of harassment and bullying. This policy should contain high-quality resources for targets to use as they decide whether or not to make a formal report to the AAS. To this end, we recommend that the AAS should also consider creating the role of an Ombudsperson(s), perhaps from within the AAS membership, and ensuring that this person(s) is appropriately trained. In addition, the AAS should create and include a brief FAQ on reporting harassment and bullying, with a link to a more detailed guide to reporting harassment and bullying (see Section 3.3). The FAQ should be easy to read and understand.

Next, it is important that the AAS continue to provide targets with multiple ways to submit reports of harassment and bullying, including by the online form, by phone, and face-to-face. After a report is made, the target should be asked if they would like to initiate an investigation. If they refuse, they may be asked if they would like their report to be saved in AAS records, with their personally identifiable information removed if they so desire.[3] If they consent to having their report saved, it may be examined by the committee should they ever be required to investigate the same perpetrator. This information repository would be confidential and potentially useful in future investigations and is not meant to serve as a resource for documenting cases of harassment and bullying for educational purposes, like the one described in recommendation 3.3. Reporting harassment and bullying to the CoEC should be distinct from the processes used to gather information on harassment and bullying for educational purposes so that targets have control over who has their information and what purpose it is being used for.

If the target requests an investigation, the committee should proceed with the steps outlined within the "The Investigation" section of the policy. Once the investigation is complete, the members of the committee who investigated the alleged instance of harassment and bullying shall make a decision regarding the application of a sanction. Any committee member who has a conflict of interest in a case should recuse themselves.

We recommend that the Code of Ethics Committee be given the authority to apply the sanctions outlined in the "Sanctions" subsection of the AAS Code of Ethics, which appears within the "Handling of Potential Ethical

Breaches" section. These sanctions include denial of privileges of AAS membership, public censure, termination of membership, and notification of sanction to home institutions. 80% of our survey respondents said removal of AAS honors, such as prizes and awards, in response to unethical behavior, would be a somewhat effective or very effective strategy to improve professional ethics in astronomy. The AAS should add to this list, the retroactive removal of distinguished honors and awards. With the exception of private reprimands as described in the AAS Code of Ethics, we encourage the AAS to make public in a transparent way any sanctions that they impose.[4]

We recommend that if a perpetrator is found by the Code of Ethics committee to have violated the AAS sexual harassment policy, the Title IX office of their home institution should *always be informed* (this currently is optional), given the consent of the target. It is vital to notify perpetrators' home institutions because the AAS's findings of harassment and bullying may validate earlier allegations against the same perpetrator, and this information may allow the home institution to pursue their own course of action. We recommend that funding agencies (NSF, NASA, DOE) should also be informed by the AAS, even if the perpetrator does not have an active grant with those organizations. The AAS should work with funding agencies to make sure that this information is appropriately disseminated to contracting organizations, e.g. the Space Telescope Science Institute (STScI).

Targets of harassment and bullying should always be in control of the extent to which their story is shared. Under no circumstances should confidentiality agreements or non-disclosure agreements be used to silence those who identify as targets of harassment and bullying, regardless of the outcomes of any investigations. Taken together, we believe these recommendations will force changes that will increase the number of targets of harassment and bullying who feel safe reporting to the AAS. Better education and a strong procedural framework for addressing these behaviors will move the culture of the AAS and of the field as a whole towards one that does not tolerate harassment and bullying.

## 3.2 Address Minor Forms of Harassment and Bullying Using a Restorative Approach

As shown by the NASEM report, minor instances of harassment and bullying can be damaging, and action is needed to curb the perpetuation of these behaviors. Many of our survey respondents expressed frustration that no mechanism is available to address minor but persistent and destructive forms of harassment and bullying. 82% of our survey respondents said that responding promptly when astronomers publicly engage in racism, sexism, ableism, and/or anti-LGBTQ+ discrimination during meetings sponsored by the AAS and its divisions would be a somewhat effective or very effective strategy to improve professional ethics in astronomy. In this section, we recommend a response strategy.

The AAS, preferably through the ODI introduced in Section 2, should create and manage an information escrow to be used to identify perpetrators of minor forms of bullying and harassment. Information escrow is a process by which parties can pass on information to a neutral third party. Once the neutral third party receives a certain amount of information, they take action. 69% of our respondents said creating an information escrow would be a somewhat effective or very effective policy to address harassment and bullying.

Meeting and event attendees can use this escrow to confidentially report minor instances of harassment and bullying, preferably through an online form. This should be a distinct resource from the one described in Section [3.1.2](), which, if created, is to be used for more serious cases of harassment and bullying, and distinct from resources maintained for educational purposes, which will be described in Section [3.3]().

As one example of how this system could work, once the neutral party receives three or more allegations against the same perpetrator through the online form, they will notify the chair of the Code of Ethics committee. The AAS office and the Code of Ethics committee should develop a policy and process on how to handle these repeat allegations. Ideally an ODI staff member trained extensively on how to respond to harassment and bullying can reach out to the perpetrator to discuss the problematic behavior. To prevent retaliation, AAS staff should not contact perpetrators during AAS meetings and events. Personally identifiable information of the submitting party should be held confidential and they should only be contacted if the Code of Ethics committee deems it warranted. In these cases, a dedicated AAS staff member should email the target with information about the resources the AAS provides for targets of more severe forms of harassment and bullying (See Section [3.3]()).

Giving perpetrators a chance to talk to an expert to learn how and why their behavior has a negative impact and how they can improve is a direct strategy the AAS can employ to halt harassment and bullying and spread knowledge about harassing behavior in order to prevent it. This proposed program can be categorized as a type of bias incident response. Many colleges and universities have some sort of bias incident response team or program in place. Universities with strong bias response programs include the University of California Santa Barbara and Oregon State University. Links to more information on these programs can be found in the references section ([Oregon State University, n.d.](); [University of California Santa Barbara, n.d.]()).

## 3.3 Create an Online Guide for Targets of Harassment and Bullying

This recommendation, and recommendations in Sections [3.4]() and [3.5](), describes actions the AAS can take to support people who are deciding whether to file a harassment and bullying complaint with the AAS. 82% of our respondents said that supporting people who are deciding whether to file a complaint with the AAS is somewhat important or very important. Many of our survey respondents asked that the AAS create a guide to navigating harassment and bullying.

We recommend that the AAS should create a detailed guide to navigating and reporting harassment and bullying and feature it on the AAS website. Such a guide should be created by communications professionals and anti-harassment and anti-bullying experts and should be tailored to the needs of AAS members. It should help people identify if they have been harassed or bullied and connect them to the proper resources for making confidential or official reports. A informative guide will help people determine the aspects of a possibly toxic work environment or encounter that may constitute harassment and bullying by providing clear descriptions of the ways harassment and bullying manifests, including the three types of sexual harassment (Sec. [3.1.1](#)) described by the NASEM report. It will also help those affected by harassment and bullying decide on a course of action by explaining how a report to their institution or the AAS is likely to proceed, and the potential disciplinary consequences that may result from an investigation. This guide will also explain the potential mental health consequences that may be experienced by the submitting party and how these consequences can be mitigated.

Our survey respondents expressed that one of the greatest challenges to reducing harassment and bullying in astronomy is a lack of understanding of how harassment and bullying manifests. Real world examples are an important tool that the AAS can use to bring harassment and bullying out of the abstract and raise the stakes of the issue. To that end, the AAS should curate a set of anonymized accounts of astronomers' experiences with harassment and bullying to publish in this resource guide. These instances should only be drawn from those reported to the Code of Ethics Committee. Because of the sensitive nature of harassment and bullying, the AAS should employ an expert who is able to solicit accounts of harassment and bullying while putting the needs of targets first and ensuring that all identities remain confidential. No account should be published or disseminated without explicit permission of the submitting party.

## 3.4 Employ an Advocate to Support Targets of harassment and Bullying

Colleges, universities, and other institutions commonly employ advocates. These advocates work directly with survivors of sexual violence to guide them through reporting processes and help them obtain other services, such as mental healthcare. They act as an intermediary between those who have been targets of sexual violence and the systems of justice that address sexual violence. These advocates usually hold a Master's in Social Work or an equivalent degree and are trained to navigate the sensitive environments around accusations of sexual violence, including harassment and bullying [(U.S. Department of Justice, 2017)](#).

73% of our survey respondents rated providing mentoring and/or counseling for those who have been adversely impacted by harassment and bullying as an important or very important way the AAS can support those who have been harmed by harassment and bullying. A complete harassment and bullying reporting, investigation, and response system should include an advocate with the appropriate qualifications. We recommend that the AAS should employ a trained, experienced advocate to work as a member of the Executive Officer's staff, preferably within an ODI. For the purposes of this recommendation, an advocate employed by the AAS should be available to support any astronomer who has experienced any form of harassment or bullying. Among other duties, this advocate should directly support targets in deciding whether or not to file a complaint, advise as to whom they should file (their institution, the AAS, or someone else), and should provide relevant support to the targets throughout the investigation process.

In addition to advocacy work, an advocate can help administer many of the anti-harassment and anti-bullying programs we recommend. The advocate, in the role as an anti-harassment and anti-bullying expert within the proposed ODI, can manage the bias response program outlined in Section 3.2. They can also create and edit some of the content of the online guide described in Section 3.3 and may be the best person to gather and publish true accounts of harassment and bullying, also as described in Section 3.3. In general, the AAS will benefit from an internal anti-harassment and anti-bullying expert who can apply best practices to the AAS's anti-harassment and anti-bullying programs.

## 3.5 Create, Administer, and Support Astronomy-Specific Anti-harassment and Anti-bullying Training

Eradicating harassment and bullying involves cultural change, and institutions often address this need through trainings to increase awareness of what harassment and bullying looks like, how it can be prevented, and the resources available to targets of harassment and bullying. Among institutions, the content, length, and form of trainings are non-standardized. The following recommendations address the need for better education to prevent harassment and bullying. We recommend that the AAS develop astronomy-specific anti-harassment and anti-bullying trainings, administer them during meetings, and share astronomy-specific anti-harassment and anti-bullying training materials among institutions.

In our survey, we asked astronomers to rate the effectiveness of different types of anti-harassment and anti-bullying trainings and to describe their experiences with trainings. From these data, we gathered a list of six traits of effective trainings that the AAS should use when developing its anti-harassment and anti-bullying training curriculum.

1. *Bystander intervention should always be taught as a core component of anti-harassment and anti-bullying training.* 74% of respondents to our survey said utilizing bystander intervention training would be a "somewhat effective" or "very effective" strategy to reduce harassment and bullying.

2. *Trainings should be developed by experts and administered by training professionals.* Our survey respondents emphasized that harassment and bullying cannot be addressed without proper professional support.

3. *Content should be specific to the power structures present in the workplaces of the target audience.* Several of our survey respondents felt that the anti-harassment and anti-bullying trainings available to them were not appropriate for physical science workplaces or academic astronomy.

4. *Trainings should demonstrate how the possession of more than one marginalized identity may lead to increased experiences of harassment and bullying.* Our survey respondents stressed the importance of increasing awareness of the harassment and bullying experienced by those with intersectional identities.

> *Trainings should center around active engagement.* Astronomers should be asked to put the ideas presented into practice through examples and scenarios, and to think about, and possibly share, the relevance of materials to their own experiences.

6. *Trainings should include a balanced combination of anonymized, real-life examples of harassment and bullying and discrimination and reputable data on prevalence and impact.* Scientific audiences are more likely to buy into a narrative supported by both quantitative and qualitative data.

Using these six traits, AAS staff should develop an anti-harassment and anti-bullying training curriculum to be deployed in various settings, examples of which are given in the following subsubsections.

### 3.5.1 Require Attendees of AAS-sponsored Meetings and Events to Engage with Anti-harassment and Anti-bullying Trainings

AAS meetings bring together thousands of people from all over the world. The AAS must do more to facilitate a shared understanding of harassment and bullying among this diverse group. At present, AAS members are required to read and agree with the Anti-harassment and Anti-bullying Policy for AAS & Divisional Meetings & Activities (Marvel, 2019). While this is an important requirement, it is insufficient because there is nothing to stop members from ignoring the text, clicking a box, and moving on.

Ideally, all AAS members and meeting attendees would attend a high-quality anti-harassment and anti-bullying training at their home institution every year. Although the AAS can take steps to promote the implementation of excellent anti-harassment and anti-bullying training at all institutions, which we will describe in Section 3.5.2, the AAS cannot guarantee training will reach every member and meeting attendee prior to AAS meetings and events. In order to increase shared knowledge about harassment and bullying, we recommend that the AAS require all members and meeting attendees, including regional and division meeting attendees, to complete a short online interactive training (e.g., a video that addresses multiple types of harassment and bullying, as well as mental health, physical health, and intellectual impacts of harassment and bullying) before registration and requires attendees to answer a small number of questions regarding common misconceptions about harassment and bullying. Such a training or accompanying short video will improve the understanding of harassment and bullying among AAS members and meeting attendees. It should be updated along with updates to the AAS's definitions of harassment and bullying.

We also recommend that the AAS should provide in-person and more in-depth anti-harassment and anti-bullying and bystander trainings at AAS meetings. These trainings should occur 3-5 times during each AAS meeting to allow all astronomers a chance to attend. Trainings can also be made available as a session at regional and division meetings. The AAS should advertise their trainings prior to meetings and encourage attendees to add it to their schedules. AAS leaders should send out notifications encouraging AAS members to attend trainings through multiple avenues (such as email, social media, and the AAS meeting app). Similar trainings have been occasionally offered at other peer professional conferences, e.g. the Lunar and Planetary Science Conference.

### 3.5.2 Provide trainings to institutions

In addition to offering training at AAS meetings, we recommend that the AAS should publish, maintain, and publicize a list of vetted training programs that are relevant to issues facing Astronomy & Physics departments. These may be offered by third-party vendors that can be contracted by individual departments. One guiding principle should be that any live (both in-person and online) training should *not* be recorded, as participants are less likely to actively engage and ask tough questions if they are being recorded.

AAS staff should regularly collect feedback from training attendees as evidence of the quality of the training. Most universities already have anti-harassment and anti-bullying trainings available to departments and groups. However, our survey respondents reported that the training that they receive at their institutions does little to illuminate the ways harassment and bullying may manifest in physical science workplaces. The AAS should reach out to departments, observatories, and other astronomical institutions and offer to share their training materials with the anti-harassment and anti-bullying experts who are internal to those institutions (such as Title IX staff members), and help them integrate astronomy-specific information into their trainings for astronomy departments.

## 3.6 Summary: Recommendations to Reduce Harassment and Bullying

One of the most powerful actions that the AAS can take to reduce harassment and bullying in astronomy is to gather and share information. The AAS should constantly collect and utilize feedback from astronomers and from field experts as it works to improve its policies and procedures. Additionally, the AAS should proactively share information about the prevalence of harassment and bullying in the sciences, the forms harassment and bullying takes, and the resources available to those who have experienced harassment and bullying. We acknowledge that, taken as a set, these recommendations are resource-intensive. A full commitment to ending harassment and bullying requires time, attention, and funding. The actions we recommend will be most effective when implemented together. Even so, should the AAS decide to pursue only some of these recommendations, we emphasize the importance of consulting with experts in the field of harassment and bullying prevention.

## 4. Recommendations to Advance Career Development

Progress towards gender parity in astronomy is slowing at every career stage, but see counterexamples from individual institutions (e.g. [Kewley et al. 2023](#)). Our survey responses demonstrate, and the literature shows, that discrimination, most commonly in the form of implicit biases, decreases women's chances of being published, receiving telescope time, and winning awards (see Paper I). As we describe in this section, the AAS should take action to mitigate the impacts of implicit biases on key processes, such as journal publication and award selection processes; we describe some of these recommended actions in detail in this section.

Academic funding structures create conditions of unequal opportunity. Many of our survey respondents reported that the academic environment is unforgiving to scientists dealing with health issues, those from low-income backgrounds, and anyone with family caregiving responsibilities. The AAS should create a compensation database and promote the use of transparent, fully informative advertisements in the AAS Job Register. These will help hold institutions accountable for providing adequate and equitable compensation.[5]

Mentorship is vital and undersupplied. As described in Section 4.8, the AAS should critically evaluate its participation in existing mentoring programs and/or create new ones if needed. Taken together, the actions recommended in this section will create more equitable conditions for all astronomers.

## 4.1 Partner with Social Scientists to Research the State of the Profession

Knowledge is a foundation of continuous improvement. Although there are some peer-reviewed studies and informative non-academic sources that reveal essential information about the conditions impacting astronomers' careers, central issues remain under-researched. All of the actions that the AAS can take to advance career development for women in astronomy can be supported by continuous research, evaluation, and revision. Presently, women and members of historically marginalized groups are often tasked with community service related to diversity, and when they cannot provide quantitative data to support their concerns, they are forced to gather it themselves. On top of their normal workplace responsibilities, women and members of historically marginalized groups should not be expected to keep track of the state of diversity in their field or subfield. We ask that the AAS leverage the ODI that we propose above to work with relevant AAS subcommittees to track the state of the field. As part of these efforts, it is important for the AAS to ensure that work does not fall disproportionately, and in an uncompensated way, on members of historically marginalized groups.

We know that poor and unequal compensation, the two-body problem, and a variety of stigmas impact astronomers and their career paths. The series of articles arising from the Longitudinal Survey of Astronomy Graduate Students (LSAGS) provide a base of research on career paths in astronomy (White, Ivie, and Chu, 2017; Porter & Ivie, 2019). However, more study is needed to build on LSAGS and address emerging and understudied issues.

We know that research partnerships between astronomers and social scientists, such as the LSAGS, the work done by Richey et al. (2019) on the risks of harassment and bullying faced by sexual and gender minorities in astronomy, and the survey work carried out by the AAS Demographics Committee are productive and informative. We recommend that the AAS ODI, in coordination with the diversity committees and the Demographics Committee, should support and expand partnerships with social scientists and other experts to conduct studies that track and examine the factors that affect astronomers' career paths into the 2020s and beyond. We recommend that the AAS should take care to ensure that the experiences of gender nonbinary scientists are accounted for in all studies, in addition to the experiences of women and men (Rasmussen et al., 2019, 2023). Key research areas for such a study include, but are not limited to, the following:

1. Career trajectories: How do factors such as the institution in which a person experienced different career stages (i.e., undergraduate, graduate, postdoctoral, faculty/staff), number of postdoctoral position(s) held, length of time spent in postdoc position(s), research area, research impact (e.g., citation metric), number and type of publications, nationality, race, gender, and sexual identity correlate with astronomers' career paths? What are the most common alternative careers to academia, and what factors influence entry into alternative careers?

2. Stigma: What do astronomers believe about themselves and other astronomers? How do these beliefs lead to stereotype threat and implicit biases, and what are the impacts of these biases? What can be done to reduce harmful stigmas, both those that come from external sources and those which originate within individuals themselves?

3. Compensation: How are astronomers compensated at each career stage, from graduate student status through the highest attainable rank for a given career path? What is the minimum acceptable annual compensation package for an astronomer at each career stage, considering variations in cost-of-living across locations and factoring in the time evolution of expenses from inflation and rising insurance costs? What maternity/caregiver leave policies should be implemented at each career stage such that the career impacts of these duties are minimized? How can institutions that provide inadequate compensation be motivated to change for the better? How can under-resourced institutions be better supported and their staff fairly compensated?

4. The two-body problem: What is the prevalence of the two-body problem, how does the two-body problem impact astronomers' career decisions and outcomes, and how has this changed over time and in response to policy changes? What can hiring institutions do to support dual-career couples?

A robust understanding of each of these topics (and others) will better allow the AAS to build on its policies and programs for supporting astronomers, especially women and members of other historically marginalized groups.

## 4.2 Model Equitable Selection Processes for Committee Membership and Awards

Science at every level, from undergraduate education through to full-professor status, is shaped by selection. Every scientific institution, including the AAS and its divisions, selects dozens of scientists each year to participate in committees and win awards. The AAS can ensure that women and members of historically marginalized groups have an equal opportunity to take on leadership roles within the AAS, and win AAS prizes, by critically examining and improving its nomination and selection processes. We recommend that the AAS should evaluate and revise its nomination and selection processes every two years using a similar process to the framework detailed in Section [3.1](#) for reviewing and revising anti-harassment and anti-bullying policies and procedures. Additionally, the AAS can encourage other astronomical institutions to follow their example and improve their own processes.

Strong nomination and selection processes will take steps to mitigate implicit bias. Examples of best practices for mitigating implicit bias include, but are not limited to, requiring selection committees to undergo implicit bias training that is specific to the task at hand and using explicit selection criteria. Candidates should be evaluated anonymously to reduce the impacts of implicit bias. Reviewers should refer to candidates using neutral terms, such as "the applicant" or "the nominee." The prize committees should be instructed to exclude criteria in STEM which are known to be biased against members of historically marginalized groups, e.g., citation indices and numbers of invited talks (e.g., [Lerman et al. 2022](#); [Chatterjee & Werner, 2021a](#),[b](#); [Nittrouer et al. 2018](#); [Schroeder et al., 2013](#)). Additionally, community service should always be included as an evaluation criteria for leadership positions and prizes, as women and members of historically marginalized groups are more likely to be asked to dedicate time to community service compared to white men. 78% of our survey respondents said that enabling greater diversity in the pool of nominees for AAS and Division prizes and committees, and capturing the data on the pools as a function of time, would be a somewhat effective or very effective policy to advance career development for women in astronomy. We recommend that the AAS should track and publish demographic data on its pools of nominees as a strategy to hold itself publicly accountable for equitable selection. The Space Telescope Science Institute's selection practices and the best practices published by the Association for Women in Science (AWIS) provide excellent bases for improving selection ([Aloisi & Reid, 2021](#); [AWIS, 2014](#)).

In our survey, 82% of the respondents said that deliberately reaching out to and involving individuals from across the entire astronomical community, especially historically marginalized and under-resourced researchers and institutions, in policy and leadership roles would be a somewhat effective or very effective policy to promote inclusion in astronomy. We recommend that the AAS should create a prize nomination committee, potentially affiliated with the ODI, whose charge should include reaching out to institutional and departmental leaders as well as faculty at minority-serving institutions (including community colleges and primarily undergraduate institutions) and those astronomy and/or physics departments with a high percentage of students and professors from groups who have been historically marginalized. To the largest extent possible, this determination should include institutional demographic/classification data that is already available[6]. This committee should encourage individuals at these institutions to nominate others or self-nominate. The committee should also consider how to increase nominations of individuals at these institutions by people outside of those institutions, perhaps by encouraging networking opportunities with past award recipients.

Once the AAS has implemented best practices for selection, we recommend that the AAS staff should publicly advertise these practices and make themselves available to consult with institutions who want to take action to improve their selection processes. We also recommend that the prize nomination committee, perhaps in cooperation with a consultant, should create a nomination toolkit to guide nominators (including self-nominators) on how to assemble a competitive nomination packet. This toolkit might include webinars and lists of recommendations for a competitive packet, including advice for letter writers to refrain from including statistics about journal impact factors or number of invited talks, unless these statistics highlight the work of women and other historically marginalized scientists (e.g., [Lerman et al. 2022](#); [Chatterjee & Werner, 2021a,b](#); [Nittrouer et al. 2018](#); [Schroeder et al., 2013](#)). We further recommend that the AAS emulate the practice of the prize committees of the Division for Planetary Sciences (DPS) and provide substantive explanations of the criteria, selection policies, and procedures for each award. This will be especially useful to guide letter writers on what characteristics of the applicants to highlight in their letters and what elements of their CV nominees should include. Finally, this committee should gather information on the demographics of prize nominees and recipients to regularly assess whether its efforts have resulted in a diversification of the prize recipients and to recommend changes and improvements if needed. By implementing equitable nomination and selection practices, the AAS can have an impact on equity both within itself and throughout the profession.

## 4.3 Improve AAS Journal Review Processes

Recent studies ([Basson et al., 2023](#); [Squazzoni et al., 2021](#)) reported that submission data for high-level journals indicate that women are less likely than men to submit to these journals and less likely to deem their work as worthy of submission to these journals, and that these differences were exacerbated by the COVID-19 pandemic. When authors and reviewers are unknown to each other - a process known as dual-anonymous review - studies (e.g., [Fox et al., 2023](#)), indicate that dual-anonymous peer review reduces reviewer bias, though no evidence was presented for a reduction in bias with respect to gender. Several journals, however, are moving to dual-anonymous peer review in an effort to mitigate gender biases. These include Elsevier and IOP Publishing. However, as we describe below, there is ample evidence in Astronomy for the benefit of dual-anonymous peer reviews in the context of the proposal review process. Motivated by these data, and under the assumption that beneficial practices in proposal review will translate to progress in mitigating journal submission and acceptance inequities, we make two recommendations to improve the AAS journal submission and review processes.

### 4.3.1 Require Dual-Anonymous Review

Single-anonymous review processes, in which the identity of the author is known to the reviewers, is found to favor famous authors and authors from prestigious institutions, disadvantaging early-career researchers and researchers from less prestigious institutions, who, in astronomy, are more likely to be women ([NCSES, 2018](#); [Tomkins et al., 2017](#)). According to the journal publishing company Wiley, dual-anonymous review is the most common form of review among social science and humanities journals, while single-anonymous review is the most common form of review among science journals ([Wiley Author Services, n.d.](#)). We recommend that the AAS journals should lead the physical sciences in implementing dual-anonymous review for journal articles. Although in small fields like astronomy, the process of removing personally identifiable information from a document does not always guarantee that an author's identity will be unknown to the reviewer(s), it is worth implementing because using neutral, anonymous language is found to support equity even when the identity of the author is known to the reviewer [(Aloisi & Reid, 2021)](#). 72% of our survey respondents rated making dual-anonymous refereeing of papers mandatory for AAS journals as a little effective, effective, or very effective as a strategy to advance professional development for women in astronomy.

While this section refers to the review process for journal articles, the best examples we currently have for the efficacy of dual-anonymous review comes from how it is implemented in the proposal review process. For example, in 2018, the Space Telescope Science Institute implemented a dual-anonymous review process for Hubble telescope time. When this type of review was implemented, proposals from women and men principal investigators (PIs) were equally as likely to be selected compared to previous cycles, in which proposals led by women PIs were less likely to be selected ([Aloisi & Reid, 2021](#); [Lonsdale et al., 2016](#); [Patat, 2016](#); [Reid, 2014](#)). The case of Hubble telescope time reveals that switching to dual-anonymous review is possible in astronomy and that it can be an effective method to advance equity in review processes.

### 4.3.2 Systematically Collect and Release Article Submission and Acceptance Data

We recommend that the AAS systematically collect and publish article submission and acceptance data for its journals, including the rates at which each journal receives and accepts articles from women and historically marginalized groups compared to men and white authors. By collecting these data and by making them publicly available, the AAS can and should hold itself accountable for upholding an equitable review process.

## 4.4 Develop a Compensation Database with a Focus on Graduate and Postdoctoral Compensation

As we reviewed in Paper I, pay inequities exist between men and women in astronomy, and, across the board, graduate students and postdoctoral researchers are not adequately compensated. 76% of our survey respondents said that developing a salary database to support efforts towards equity in pay would be a somewhat effective or very effective strategy to create a more inclusive astronomical community. In their written responses, our survey respondents indicated that the greatest barrier to entering graduate school and persisting is compensation in early career stages. Compensation includes salary, as well as benefits such as health care, paid leave and other resources for caregivers.

Presently, informal resources, such as the Astronomy Rumor Mill, exist for reporting salaries. However, these data are unreliable. We recommend that the AAS create an official compensation database and prioritize gathering data on compensation for graduate students and postdoctoral researchers. The database should allow astronomers to anonymously report their compensation packages, their job title, their institution, and optionally, their demographic characteristics, so that these data may be listed in the database. Demographic characteristics are important, but, to encourage reporting, they should be made optional, as revealing such information may eliminate anonymity for some scientists. To the extent possible, an AAS compensation database should include data on gender differences in compensation. The AAS may encourage astronomers to collaborate with the project on social media, through email, on the AAS website, and at AAS meetings.

Increasing transparency about compensation will incentivize prospective graduate students and job applicants to more strategically negotiate their compensation packages and, consequently, motivate institutions to improve their pay and benefits, especially for graduate students and postdocs. Having such a resource available will help prospective graduate students make fully informed decisions about whether they want to pursue a career in astronomy. In addition, the AAS can use these data to publicly recognize institutions who improve compensation and feature them on the AAS website and on the Job Register.

## 4.5 Take Steps to Improve the Graduate and Postdoctoral Experience

Many of our survey respondents prioritized improving the experiences of graduate students and postdoctoral researchers as a strategy to advance the participation of women and other members of historically marginalized groups in astronomy. Graduate students and postdocs are often, but not always, in their 20s and 30s. At this stage of life, many people begin to "settle down," as they cement long-term partnerships, seek to establish financial security, and look towards starting families. Compensation packages for graduate students are inconsistent across institutions, and often insufficient. Postdoctoral positions usually last three years and often do not pay at a rate that is commensurate with experience. Additionally, they often require astronomers to move to one location for a three-year period, and then uproot themselves to move on to a new position.

Many astronomers, but especially women and other members of historically marginalized groups, suffer in postdoctoral positions because of the low compensation and lack of long-term stability offered by postdoctoral work. If graduate and postdoctoral work continue to be typical career stages in astronomy, then these career stages must provide the robust support that early-career researchers need to thrive in early adult life.

The Astro2020 white paper *The Early Career Perspective on the Coming Decade, Astrophysics Career Paths, and the Decadal Survey Process* recommends that postdoctoral programs be improved by lengthening the time of postdoctoral positions, instituting formal mentorship programs, providing pathways to long-term positions at the same institution, and guaranteeing healthcare benefits and caregiver leave [(Moravec et al., 2019)](). This white paper can serve as a basis for initiating discussion on the issues early-career researchers face. For example, we recommend that the AAS host sessions at AAS meetings for astronomers to convene to discuss the challenges of early-career employment. We also recommend that the AAS then address the issues raised at these sessions by creating resources that departments can use when designing and revising their graduate and postdoctoral programs. For example, the AAS may consider creating a guide to help departments advocate for institutional maternity and caregiver policies that are ideal for each career stage. The experience of being a caregiver as a postdoc and as a graduate student likely differ and may require different policies and resources. Finally, we recommend that the AAS encourage grant PIs to offer paid leave (within institutional guidelines) for all graduate students and postdocs. These steps are important to avoid putting these researchers in a financially difficult position. The AAS can play a key role in improving early-career employment by investigating and promoting best practices for PIs to support these early-career scientists.

## 4.6 Address the Two-Body Problem

Astronomers at all career stages commonly face the two-body problem, in which both partners are astronomers and/or both are specialized enough in their respective fields that one would not find employment in the same general location as the other. Many of our respondents expressed relief that the issue of the two-body problem was included in our survey, as they felt that much of the existing discussion and action (or lack thereof) around the issue did not provide any meaningful outcomes for them.

The two-body problem is highly stigmatized. Our survey respondents reported that often, when those facing the two-body problem seek advice and guidance, they are counseled to simply choose between career and family, or seek a new partner. Just as the AAS should facilitate dialogue on the issues impacting early-career scientists, we recommend that the AAS facilitate dialogue on the two-body problem at AAS meetings and sponsored events. Publicly acknowledging the widespread impact of the two-body problem may lift some of the stigma associated with it.

Our survey revealed two strategies for coping with the two-body problem that are relatively popular among astronomers: negotiating for a second position and enabling remote work. Of our survey respondents who have experienced the two-body problem, only 44% found negotiating for a shared position (i.e., one position shared by two people) to be effective, whereas 78% found negotiating for a second position to be effective. In addition, 83% of our respondents rated enabling remote work as a somewhat important, important, or very important strategy to alleviate the two-body problem. As the COVID-19 pandemic has necessitated and normalized remote work, it may be more feasible than ever for institutions to offer remote work flexibility as part of their employment packages. We recommend that the AAS seize on this opportunity and encourage its membership to implement the lessons learned from remote work during the COVID-19 global pandemic to integrate more remote work into their strategies for assisting dual-career couples in the long-term.

To help AAS members advocate for their institutions to adopt excellent, detailed, and transparent policies for assisting dual-career couples, we recommend that the AAS create and distribute a best-practices guide on the two-body problem. The CSWA has a large set of [resources](#) on its AAS web page and a series of [Women in Astronomy Blog](#) posts from department chairs who discussed their ability to find positions for spouses. This guide should take into account that different institutions may need different solutions: smaller, under-resourced institutions may be unable to create second positions, while larger institutions may have more flexibility to create second positions. It should also emphasize the importance of buy-in by departmental and institutional leadership, as deans and department heads have the most control over the hiring process. This guide should be published on the AAS website and distributed to AAS members, departments, and institutions. AAS staff should make themselves available to consult with members as well as with department and institutional leadership who want to take action to better support dual-career couples. Best practices for dealing with the two-body problem can be adapted from excellent publicly available resources, such as the *Support for Dual Career Couples* guide from the StratEGIC Toolkit for Effecting Gender Equity and Institutional Change [(Laursen & Austin, 2014)](). A further list of resources on the two-body problem, compiled by the CSWA, is available on the AAS's website [(CSWA, 2013)]().

## 4.7 Continue to Educate Astronomers About Alternative Careers

Our survey respondents revealed that there is an unmet demand in astronomy to educate graduate students and others about career alternatives. For example, as graduate students work to finish their degrees, they may realize that a career as a full-time academic is not the best fit. They often have limited access to mentors who have faced the same decision point and may not feel comfortable sharing their thoughts about leaving academia with their academic peers and mentors. There are opportunities for people who were trained as astronomers but who have experience working in non-academic fields, including science journalism, data science, K-12 education, and other industry or government positions to provide information about careers in these fields and the tradeoffs compared to careers in academia.

We recommend that the AAS work with both academic and non-academic partners to create and promote more educational materials about career alternatives. At the present, the AAS features some non-academic career profiles on its website and, while this is an excellent start, there is potential to build on this content. The AAS can publish news items about astronomers who take alternate career paths and should continue to host panels at AAS meetings and as virtual webinars as a venue for PhD astronomers in non-traditional positions to talk about their experiences. The AAS has already hosted these kinds of events, e.g., the series of Career Sessions on *Astronomers Turned Data Scientists* (AAS 237, AAS 243). The AAS and its members must normalize transitions to and from alternate career paths, first and foremost so that astronomers can make the best choices for their future; second, so that the astronomical community builds healthy relationships with non-academic institutions; and third, so that individuals feel welcome to pursue more traditional astronomical career paths in the future if they so desire.

## 4.8 Support a Distance Mentorship Program

Mentorship is often a critical factor that shapes the career path of a scientist. Ideally, all astronomers would have excellent mentors within their institutions, but we know this is not the case, especially for women and other members of historically marginalized groups. The LSAGS found that women in graduate astronomy programs were more likely than their male peers to seek mentorship outside of their primary advisor, demonstrating that women often find themselves in need of supplementary mentorship [(White, Ivie, and Chu,]() 2017; Porter & Ivie, 2019). Additionally, a recent study from the AIP National Task Force to Elevate African American Representation in Undergraduate Physics & Astronomy found that undergraduate women in physics and astronomy are less likely to be encouraged and affirmed by their professors, display lower levels of classroom self-efficacy compared to men, and are slightly less likely to perceive themselves as physicists (James et al., 2020). The AAS can and should take action to close these gaps by investing more in mentorship.

In our survey, 77% of respondents said offering a mentoring program for astronomers at all career stages would be a somewhat effective or very effective policy to advance professional development. 73% rated providing counseling/mentoring for those who have been adversely affected by harassment and bullying as somewhat important or very important. 55% of our survey respondents rated finding ways to encourage and mentor those dealing with the two-body problem as important or very important. 84% of our survey respondents rated it as an effective or very effective strategy to provide support for astronomers at all career stages when those astronomers are from small institutions or non-academic organizations who may not have access to the same support network as those at larger institutions.

These data reveal that there are mentorship needs in astronomy that are unfulfilled. We recommend that the AAS work with professionals with expertise in mentoring within the physical sciences to develop an effective mentoring program and to recruit astronomers from both academic and non-academic institutions to serve as mentors[7]. Mentors should undergo training on best practices in mentoring, anti-harassment and anti-bullying, and mitigating unconscious bias. Mentees who apply to the program should similarly be asked to describe their needs so that the AAS can use this information to match mentees to mentors. Within the limits of mandatory reporting guidelines, this information should remain confidential. The AAS should recruit mentors and mentees from its membership and should also allow participation by mentees who are non-members so that those who cannot afford AAS dues have an opportunity to participate. Members of historically marginalized groups can be valuable mentors but are already overburdened with official and unrecognized service. It is therefore critical that mentors are compensated for their AAS mentoring duties. A mentorship program offered by the AAS will act as one more service provider in the market of mentorship opportunities available to astronomers.

## 4.9 Summary: Recommendations to Advance Career Development

We emphasize the AAS's power as an advocate for systemic change in the sciences. The AAS should push universities and federal funding agencies to treat the people performing science as central to excellent scientific outcomes. We call on the AAS to continue to support and acknowledge the activities of other groups internal to AAS that strive to improve the professional climate in the field, including the other diversity committees, the Site Visit Oversight Committee, the Demographics Committee, the Education Committee, the Committee on Employment, and others. The quality of life of scientists is *key to creating quality science.* Additionally, the AAS is also part of a larger scientific system that is at risk if action is not taken to redesign science to be a viable career path for anyone and everyone who has the ability and drive to work in science. In addition to providing programs and information to help promote excellent pathways at all career stages, the AAS should partner with other professional scientific societies, such as APS and AGU, to advocate widely for a more equitable and sustainable scientific professional system.

We believe the AAS should revise and expand its resources to support astronomers at all career stages, setting an example and acting as a consultant to help departments and their institutions implement equitable practices. AAS staff should carefully coordinate their programs with one another and with the AAS committees and remain knowledgeable about the AAS's full range of resources in order to refer astronomers to the service that best fits their needs. We acknowledge that many of these programs and resources will take time to put in place. However, in the long term, as these recommendations are implemented, evaluated, and improved, they will support progress towards lifting many of the barriers to success faced by women in astronomy, members of other historically marginalized groups, and indeed all members of the astronomical community.

A topic not specifically addressed by our survey is how improvements to the AAS Job Register may maximize its effectiveness in attracting job applications from individuals belonging to historically marginalized groups. While we have no specific suggestions in this area, we encourage the AAS and the Employment Committee to continue to address this issue.

## 5. Recommendations to Continuously Improve AAS Meetings

AAS meetings are the most important professional development opportunity that the AAS provides for the astronomical community. Therefore, the AAS has a responsibility to remove barriers to meeting participation to ensure that all astronomers are able to engage in all of the activities that take place at meetings. The AAS should work continuously to make AAS meetings as accessible and inclusive as possible.

To demonstrate its commitment to diversity and inclusion and to initiate important conversations among diverse groups of astronomers, the AAS must continue to support meeting sessions on the issues that impact women and other members of historically marginalized groups in astronomy. In this section, we propose a framework to help the AAS continuously collect feedback from astronomers about their experiences at AAS meetings, paying special attention to members of historically marginalized groups. We also voice our support for meeting sessions to discuss topics such as leading diverse teams, the two-body problem, harassment and bullying. Taken together, these actions will improve the meeting experiences of astronomers of all identity groups and have positive impacts on the future of the profession.

### 5.1 Continue and Improve Efforts to Advance Inclusivity at AAS Meetings

The AAS needs to normalize inclusivity at all meetings organized by or sponsored by the AAS including national meetings, regional meetings, conferences, webinars, virtual meetings, or other events. An inclusive AAS meeting will serve astronomers across all axes of diversity. 73% of our survey respondents said that working with SGMA, WGAD, CSMA, CSWA, and others to promote gender-neutral bathrooms, lactation rooms, and other provisions[8] for marginalized groups would be a somewhat effective or very effective strategy to increase inclusivity. Moreover, 77% of our survey respondents indicated that providing funding that increases the accessibility of networking and meeting spaces would be somewhat effective or or very effective at fostering equity and inclusion, especially for community members with intersectional marginalized identities. 87% of our survey respondents said that scheduling meetings at family-friendly times, being flexible[9] when scheduling events, and providing video conferencing capabilities would be a somewhat effective or very effective strategy to increase inclusivity in astronomy. 81% of our survey respondents said that, for all career levels, providing funding and access to resources to mitigate the extra impact of caregiving on women would be a somewhat effective or very effective strategy to improve professional development for women in astronomy.

It is challenging to define a uniform standard of "family-friendly times" that can be applied to a variety of venues and media. However, now that the AAS has hosted multiple meetings in an entirely digital and hybrid format, providing online access to AAS meetings is more feasible than ever and may be one way to make meetings more "family friendly". Therefore, we recommend that the AAS should continue to provide at least some online access to selected aspects of its meetings. We also recommend that funding be made readily available to increase the accessibility of networking and meeting spaces, where meetings are defined broadly as above. There should be a well-established procedure to apply for funding for this kind of activity for any AAS-sponsored meeting or conference, and this procedure should include a list of recommended inclusivity practices that is available to the applications. Stating the plans for providing accommodations should be a required part of the application process for any such meeting or conference. We suggest that the AAS work with the relevant inclusion committees to ensure that gender-neutral restrooms, lactation rooms, and other provisions are easily accessible and their availability clearly communicated to meeting attendees. The organizers of AAS-sponsored meetings and conferences should also ensure that the availability of any provisions, such as those listed above, are clearly communicated to the participants.

We will not list out more inclusive practices for meetings here. Instead, we encourage the AAS to utilize the excellent work of other groups who have created guidance on inclusive conferences, including *Inclusive Scientific Meetings: Where to Start* (Pendergrass et al. 2019); *LGBTQPAN+ Inclusivity and Physics and Astronomy: A Best Practices Guide* (Ackerman et al., 2018); WGAD's Astro2020 white paper (Aarnio et al., 2019); the "Nashville Recommendations" (Inaugural Inclusive Astronomy Meeting, 2015); and *Recommendations for Planning Inclusive Astronomy Conferences* (Local Organizing Committee, Inclusive Astronomy 2 et al., 2020).

We also recommend that the AAS commit to soliciting feedback from meeting attendees on the inclusivity of the meeting and to doing so after every AAS meeting. The AAS should keep its strategy for ensuring an inclusive meeting up-to-date and easily available. As part of this process, they should actively seek community feedback on the strategy, including feedback from all four diversity committees. For example, the AAS started providing dependent care grants and childcare services for meetings [(American Astronomical Society, n.d.)](#) after they responded to community feedback. The AAS can improve such services by proactively collecting and responding to feedback on them and by ensuring that they are accessible to astronomers at all career levels. By regularly seeking and responding to input from the astronomical community, the AAS will remain accountable for delivering the resources and services that it promises.

## 5.2 Continue to Support Diversity and Inclusion Sessions at AAS Meetings

Throughout the past decade, meeting sessions on diversity and inclusion, often sponsored and cosponsored by the diversity committees, have become a regular presence at AAS meetings. These sessions are central to the function of the diversity committees because they allow them to connect with the astronomers that we serve and to identify the issues that are important to them. 76% of our survey respondents said that continuing to provide support for sessions that focus on diversity, equity, and inclusion issues at AAS meetings would be a somewhat effective or very effective strategy the AAS can use to advance diversity and inclusion in astronomy. In order to better support these meeting sessions and allow astronomers who are interested in issues of diversity, equity, and inclusion to learn as much as possible about these issues, we recommend that the AAS should take care to avoid scheduling sessions on diversity, equity, and inclusion at overlapping times.

Our survey respondents reported that while sessions on issues of equity are important, they are often attended by the same set of people who are already knowledgeable about the issues that impact women and other members of historically marginalized groups. We recommend that the AAS leadership work to increase the visibility of these sessions. To help these sessions reach a broader audience, the AAS leadership can publicly commit to attending some of or all of the sessions on diversity and inclusion at each meeting. AAS leaders should advertise the sessions they will attend on social media and by email, and should actively encourage other senior astronomers to accompany them. Validation by leadership provides legitimacy and will attract larger groups of astronomers to these sessions. The AAS may also consider compiling a full list of meeting sessions on diversity, equity, and inclusion prior to each meeting and publicizing it on social media to increase awareness of these sessions.

## 5.3 Summary: Recommendations to Improve AAS Meetings

Our vision for the AAS is a professional society where every scientist feels personally and intellectually valued. AAS meetings are the AAS's greatest signal of its culture, ideals, and future direction. Each and every AAS meeting should be free from discrimination and harassment and bullying and should be accessible and welcoming to all astronomers.The vast majority of survey respondents indicated that ignoring the issues impacting women and other members of other historically marginalized groups during AAS meetings holds science back by silencing members of the scientific community. To make AAS meetings and other activities welcoming to all astronomers will require ongoing action by the AAS and its membership to ensure that meetings are made more inclusive and accessible. In addition, finding effective and sustainable solutions to the issues of discrimination and harassment and bullying in our community will require the combined problem solving power of astronomers (and non-astronomers) across all intersections. Sessions focused on equity, diversity, and inclusion at AAS meetings are an important forum for such discussions and should be regularly supported for the foreseeable future. These actions will help the AAS to harness the full intellectual potential of all community members.

## 6. Summary

This white paper addresses the CSWA's mandate to make practical recommendations to AAS leaders on actions to advance the status of women in astronomy. Data to support these recommendations come primarily from an original survey deployed by the CSWA in the spring of 2019 (see Paper I). In shaping our recommendations we drew not only on this survey but also on reports from the National Science Foundation, the National Academies of Science, Engineering, and Medicine (NASEM), academic literature, and white papers on diversity and inclusion written by astronomers. Our review of the literature found that progress towards gender parity between men and women in doctoral degree earning is slowing, that women are overlooked for tenure-track positions in academia, and that women tend to drop out of research at high rates. These data indicate that intervention is needed to support women as they pursue scientific excellence. The primary purpose of the CSWA survey was to determine the priorities of actions that the AAS could take and to solicit other actions that could be employed (or not), according to a broad cross-section of the AAS community. While our recommendations are well-informed by the survey and by current literature, we are not experts in the sociology of science. The successful implementation of these recommendations will require the input of experts in relevant areas outside of astronomy.

Women, especially women of color, and LGBTQ+ women scientists are harassed at high rates. The NASEM report made important progress towards ending sexual harassment in the sciences by identifying forms of sexual harassment e.g., unwanted sexual attention, sexual coercion, and gender harassment, and by revealing the impacts sexual harassment can have on both targets of harassment and on the overall climate of environments where harassment is present. Additionally, implicit biases, a lack of transparency, inadequate compensation, and a lack of support for caregivers and partnered astronomers create barriers to success. While these issues clearly impact women, more research is needed to fully understand how women's careers in astronomy are negatively affected by stigma, poor compensation, the two-body problem, and caregiving responsibilities. However, armed with the current information, the AAS can and should already take action to end harassment and bullying in astronomy by strengthening its harassment and bullying reporting and sanctioning procedures, by addressing minor forms of harassment and bullying using a restorative approach, and by creating and distributing field-specific anti-harassment and anti-bullying training materials. The AAS can also begin to mitigate the impacts of implicit bias by assessing, improving, and making transparent its selection processes for awards and committee positions. Further, the AAS journals can implement dual-anonymous peer review, and it can support equitable compensation by creating a compensation database.

Finally, the AAS can help more astronomers reach impactful mentors by critically evaluating its participation in mentoring365 and by creating a new distance mentorship program if the mentoring365 participation is not effective.

Throughout this paper, we have called on the AAS to take specific actions to advance the status of women in astronomy. The AAS can create new resources that will meet previously unaddressed needs and improve its existing resources to better serve astronomers of all identities. We take the stance that the AAS can best facilitate these initiatives by creating an Office of Diversity and Inclusion, but we urge the AAS to take on most, if not all, of these initiatives regardless of whether a new office is established. We take the stance that the astronomical community and the AAS would both benefit by having the AAS's consideration of these recommendations and their steps to address them clearly communicated to the AAS membership.

We recognize that these recommendations do not include estimates of cost and time; the AAS will have to develop a roadmap to address and implement the suggested recommendations. However, based on the AAS Strategic Plan, and especially Strategic Priorities 2 and 3 (AAS, 2021), we believe that none of our recommendations are outside of the scope of the AAS's purpose and ability.

Continuing to investigate and mitigate the causes of structural barriers to achievement for women in astronomy is just as vital today as it has been since the CSWA was established in 1979. Equity and inclusion is not a zero sum game: improving the diversity and inclusiveness of astronomy as a field will lead to high-quality, impactful science. When astronomers have no choice but to work in environments characterized by harassment and bullying and encounter many other significant barriers, they often cannot fully join in pushing astronomy forward. We look forward to the scientific achievements of a workforce that fully utilizes the talents of all its members. Although this paper was written specifically for the AAS, we invite scientists from all disciplines to review these actions and consider their applicability to other professional societies and scientific institutions. We hope that within the next decade, the AAS becomes an exemplary professional society for diversity and inclusion in the sciences.

## Acknowledgements

Wexler thanks NASA for its support through the NASA internship program. The authors thank the astronomical community for their responses to the survey, which directly enabled this work. Rudnick thanks the NSF for their support through AAG grants AST-1716690, 2206473, and 2308126.

# Appendix

The following is the full list of recommendations and sub-recommendations contained in this paper. The corresponding section is indicated in the first two columns.

| Table 2 | | |
| --- | --- | --- |
| Section | Sub-section recommendation | Brief recommendation text |
| 2 | | The AAS should establish an Office of Diversity and Inclusion (ODI) with two to three staff members, as a division of the Executive Officer's staff. |
| | 2a | The AAS should consider funding internships for aspiring social scientists and astronomers to work with the ODI and the diversity/inclusion committees on some of the initiatives and actions described in this paper, though not all of our recommendations are appropriate intern projects. |
| 3.1 | | The AAS should commit to critically assessing and updating its definitions of harassment and bullying, and its harassment and bullying reporting procedures, investigative processes, and response policies regularly (e.g., at least every five years). |
| 3.1 | | AAS staff members should conduct a full review of literature on harassment and bullying in the sciences and best practices for addressing reports of harassment and bullying, with a focus on materials published in the past five years. |
| 3.1 | | The AAS should collect the input of its membership regarding the effectiveness of its anti-harassment and anti-bullying policies. |
| 3.1.1 | | The AAS should ensure consistency in how harassment and bullying and bullying are defined in the AAS's Code of Ethics and the Anti-Harassment Policy for AAS & Division Meetings & Activities. |
| | 3.1.1a | The AAS's unified definition of harassment and bullying should include a baseline definition of harassment and bullying followed by subsections for sexual , racial, ableist, ageist, and forms of harassment and bullying that may impact LGBTQ+ scientists, including harassment and bullying on the basis of sexuality, gender identity, and gender presentation. |
| | 3.1.1b | The AAS's definition of sexual harassment should be updated to be consistent with the findings of the 2018 NASEM report, which identified three types of sexual harassment: sexual coercion, unwanted sexual attention, and gender harassment. |
| 3.1.2 | | The AAS should make changes to the content and procedures contained in the sections of the Anti-Harassment Policy for AAS & Division Meetings titled "Reporting an Incident" and "The Investigation". |
| | 3.1.2a | The AAS should investigate every claim of harassment and bullying that it receives and treat cases of harassment and bullying as seriously as it treats other forms of ethical misconduct. |
| | 3.1.2b | The AAS should engage with outside expertise to aid in the investigation(s) of instances of harassment and bullying. |

|  | 3.1.2c | The AAS should also consider creating the role of an Ombudsperson(s), perhaps from within the AAS membership, and ensuring that this person(s) is appropriately trained. |
|---|---|---|
|  | 3.1.2d | The AAS should create and include a brief FAQ on reporting harassment and bullying, with a link to a more detailed guide to reporting harassment and bullying. |
|  | 3.1.2e | The AAS should continue to provide targets with multiple ways to report harassment and bullying, including by the online form, by phone, and face-to-face. |
|  | 3.1.2f | The Code of Ethics Committee be given the authority to apply the sanctions outlined in the "Sanctions" subsection of the AAS Code of Ethics, which appears within the "Handling of Potential Ethical Breaches" section. |
|  | 3.1.2g | If a perpetrator is found by the Code of Ethics committee to have violated the AAS sexual harassment policy, the Title IX office of their home institution should *always be informed* (this currently is optional), given the consent of the target. |
|  | 3.1.2h | Funding agencies (NSF, NASA, DOE) should also be informed by the AAS, even if the perpetrator does not have an active grant with those organizations. |
| 3.2 | | The AAS, preferably through the ODI introduced in Section 2, should create and manage an information escrow to be used to identify perpetrators of minor forms of bullying and harassment. Information escrow is a process by which parties can pass on information to a neutral third party. |
| 3.3 | | The AAS should create a detailed guide to navigating and reporting harassment and bullying and feature it on the AAS website. |
| 3.4 | | The AAS should employ a trained, experienced advocate to work as a member of the Executive Officer's staff, preferably within an ODI. |
| 3.5 | | The AAS should develop astronomy-specific anti-harassment and anti-bullying trainings, administer them during meetings, and share astronomy-specific anti-harassment and anti-bullying training materials among institutions. |
| 3.5.1 | | The AAS should require all members and meeting attendees, including regional and division meeting attendees, to complete a short online interactive training (e.g., a video that addresses multiple types of harassment and bullying, as well as mental health, physical health, and intellectual impacts of harassment and bullying) before registration and requires attendees to answer a small number of questions regarding common misconceptions about harassment and bullying. |

|  | 3.5.1a | The AAS should provide anti-harassment and anti-bullying and bystander trainings at AAS meetings |
| --- | --- | --- |
| 3.5.2 |  | The AAS should publish, maintain, and publicize a list of vetted training programs that are relevant to issues facing Astronomy & Physics departments. |
| 4.1 |  | The AAS ODI, in coordination with the diversity committees and the Demographics Committee, should support and expand partnerships with social scientists and other experts to conduct studies that track and examine the factors that affect astronomers' career paths into the 2020s and beyond. |
|  | 4.1a | The AAS should take care to ensure that the experiences of gender nonbinary scientists are accounted for in all studies, in addition to the experiences of women and men. |
| 4.2 |  | We recommend that the AAS should evaluate and revise its nomination and selection processes every two years using a similar process to the framework detailed in Section 3.1 for reviewing and revising anti-harassment and anti-bullying policies and procedures. |
|  | 4.2a | The AAS should track and publish demographic data on its pools of nominees as a strategy to hold itself publicly accountable for equitable selection. |

|  | 4.2b | The AAS should create a prize nomination committee, potentially affiliated with the ODI, whose charge should include reaching out to institutional and departmental leaders as well as faculty at minority-serving institutions (including community colleges and primarily undergraduate institutions) and those astronomy and/or physics departments with a high percentage of students and professors from groups who have been historically marginalized. To the largest extent possible, this determination should include institutional demographic/classification data that is already available. This committee should encourage individuals at these institutions to nominate others or self-nominate. |
|  | 4.2c | Once the AAS has implemented best practices for selection, the AAS staff should publicly advertise these practices and make themselves available to consult with institutions who want to take action to improve their selection processes |
|  | 4.2d | The AAS prize nomination committee, perhaps in cooperation with a consultant, should create a nomination toolkit to guide nominators (including self-nominators) on how to assemble a competitive nomination packet |
|  | 4.2e | The AAS should emulate the practice of the prize committees of the Division for Planetary Sciences (DPS) and provide substantive explanations of the criteria, selection policies, and procedures for each award |
| 4.3.1 | | The AAS journals should lead the physical sciences in implementing dual-anonymous review for journal articles. |
| 4.3.2 | | The AAS should systematically collect and publish article submission and acceptance data for its journals, including the rates at which each journal receives and accepts articles from women and historically marginalized groups compared to men and white authors |
| 4.4 | | The AAS should create an official compensation database and prioritize gathering data on compensation for graduate students and postdoctoral researchers |
| 4.5 | | The AAS should host sessions at AAS meetings for astronomers to convene to discuss the challenges of early-career employment. |
| | 4.5a | The AAS should then address the issues raised at these sessions by creating resources that departments can use when designing and revising their graduate and postdoctoral programs. |
| 4.5 | | AAS should encourage grant PIs to offer paid leave (within institutional guidelines) for all graduate students and postdocs. |
| 4.6 | | The AAS should facilitate dialogue on the two-body problem at AAS meetings and sponsored events. |
| 4.6 | | The AAS should encourage its membership to implement the lessons learned from remote work during the pandemic to integrate more remote work into their strategies for assisting dual-career couples in the long-term. |
| 4.6 | | The AAS should create and distribute a best-practices guide on the two-body problem. |
| 4.7 | | The AAS should work with both academic and non-academic partners to create and promote more educational materials about career alternatives. |
| 4.8 | | The AAS should work with professionals with expertise in mentoring within the physical sciences to develop an effective mentoring program and to recruit astronomers from both academic and non-academic institutions to serve as mentors. |
| 5.1 | | AAS should continue to provide at least some online access to selected aspects of its meetings. |
| | 5.1a | Funding be made readily available to increase the accessibility of networking and meeting spaces. |
| | 5.1b | The AAS should commit to soliciting feedback from meeting attendees on the inclusivity of the meeting |
| 5.2 | | The AAS should take care to avoid scheduling sessions on diversity, equity, and inclusion at overlapping times. |
| | 5.2a | The AAS leadership should work to increase the visibility of these sessions. |

## Footnotes

1. During the development of the original survey the term "members of underrepresented groups (URM)" was commonly accepted as an appropriate term. A more recent consensus is to use "members of historically marginalized groups". We refer to the latter throughout this paper, though original survey questions and comments quoted verbatim still adopt the original term. ↩

2. We acknowledge that the AAS's sphere of influence does not extend directly into some venues, e.g. university campuses. However, we feel that the policies and actions of the AAS can inspire activities in a range of settings and even provide implicit justification for changes in policies in those venues. ↩

3. Removing personally identifiable information may have the implication that it will be difficult in the future to gain permission to modify, resolve, make use of, or retract the complaint. It may therefore be desirable for the AAS to maintain a confidential database of the source of each complaint, even if the identity of the target is removed from the complaint itself. ↩

4. The August 2023 update of the AAS Code of Ethics now includes a statement recognizing Harassment and Bullying as forms of scientific misconduct. This change now allows the sanctions referred to here to be applied to individuals who the CoE committee finds have violated the Code of Ethics. This is a welcome recent change to AAS policy. ↩

5. For example, some states, e.g. NY and CA, require salary ranges to be listed in job postings. ↩

6. An example of a resources for this kind of information is https://ncses.nsf.gov/pubs/nsb202332/institutions-in-s-e-higher-education#minority-serving-institutions ↩

7. To help mentor astronomers, the AAS has engaged with mentoring365. However, it is not clear how effective or widely known this program is and we recommend that the AAS undertake a critical review of their participation in this program and its effectiveness. ↩

8. Though not explicitly specified in the survey, these should include provisions for those with visible and invisible disabilities. ↩

9. In this context, we define flexible to mean that the AAS should be willing to adjust its previously planned times for events based on new information that highlights conflicts with the schedules of key constituents. While this clearly is infeasible for large meetings, it should be proactively considered for smaller meetings and virtual events. ↩

- Laursen, S. L., & Austin, A. E. (2014). Strategic Intervention Brief #10: Support for Dual-Career Couples. StratEGIC Toolkit: Strategies for Effecting Gender Equity and Institutional Change. Retrieved from: https://www.colorado.edu/eer/sites/default/files/attached-files/10_dualcareerbrief123015.pdf

- Lerman, K., Yu, Y., Morstatter, F., & Jay Pujara. (2022). Gendered citation patterns among the scientific elite. *Proceedings of the National Academy of Sciences*, *119*(40), e2206070119. https://doi.org/10.1073/pnas.2206070119 ↵
- Local Organizing Committee, Inclusive Astronomy 2, Brooks, B., Brooks, K., Hagen, L., Hathi, N., Hoffman, S., … Prichard, L. (2020). Recommendations for Planning Inclusive Astronomy Conferences. *arXiv E-Prints*, arXiv:2007.10970. https://doi.org/10.48550/arXiv.2007.10970 ↵
- Lonsdale, C. J., Schwab, F. R., & Hunt, G. (2016). Gender-Related Systematics in the NRAO and ALMA Proposal Review Processes. *arXiv E-Prints*, arXiv:1611.04795. https://doi.org/10.48550/arXiv.1611.04795 ↵
- Marvel, K., Parriott, J., Clark, K., Scuderi, E., Frendak, D., Steffen, J., & Fienberg, R. (2019). On the AAS: A Resource for the 2020 Decadal Survey. *Bulletin of the AAS*, *51*(7). Retrieved from https://baas.aas.org/pub/2020n7i275

- Moravec, E., Czekala, I., Follette, K., Alpasian, M., Amon, A., Armentrout, W., … Youngblood, A. (2019). The Early Career Perspective on the Coming Decade, Astrophysics Career Paths, and the Decadal Survey Process. *Bulletin of the AAS*, *51*(7). Retrieved from https://baas.aas.org/pub/2020n7i008

- National Academies of Sciences, Engineering, & Medicine. (2018). *Sexual Harassment of Women: Climate, Culture, and Consequences in Academic Sciences, Engineering, and Medicine* (P. A. Johnson, S. E. Widnall, & F. F. Benya, Eds.). Washington, DC: The National Academies Press. https://doi.org/10.17226/24994 ↵
- National Center for Science and Engineering Statistics (NCSES). 2018. *Doctorate Recipients from U.S. Universities: 2017*. Special Report NSF 19-301. Alexandria, VA. Available at **https://ncses.nsf.gov/pubs/nsf19301/**

- Nittrouer, C. L., Hebl, M. R., Ashburn-Nardo, L., Trump-Steele, R. C. E., Lane, D. M., & Virginia Valian. (2018). Gender disparities in colloquium speakers at top universities. *Proceedings of the National Academy of Sciences*, *115*(1), 104–108. https://doi.org/10.1073/pnas.1708414115 ↵
- Oregon State University. n.d. Bias Incident Response Process. Retrieved from: https://diversity.oregonstate.edu/bias-incident-response-process.